\documentstyle[prb,aps,multicol,epsf]{revtex} \tighten
\begin{document}
\draft

\title{Strong tunneling and charge quantization in S-I-N Coulomb blockade
structures}

\author{K.A. Matveev$^{(1)}$ and L.I. Glazman$^{(2)}$}

\address{$^{(1)}$Department of Physics, Duke University, Durham, NC
  27708-0305\\ $^{(2)}$Theoretical Physics Institute, University of
  Minnesota, Minneapolis MN 55455}

\maketitle

\begin{abstract} 
We study the charge of a small normal island connected by a tunnel  junction
to a superconducting lead.  Unlike the N-I-N case, the steps of the Coulomb
staircase remain sharp even if the conductance of the tunneling barrier
exceeds $e^2/h$. One can observe the transition from sharp steps to the
smeared ones by applying magnetic field to destroy the superconductivity.
\end{abstract}

\begin{multicols}{2}
  
The charge of a small conducting island weakly coupled to an electron
reservoir (lead) by a tunneling barrier shows a step-like dependence on
the gate voltage due to the well-known phenomenon of Coulomb
blockade.\cite{GrabDev} It was shown in
Refs.~\onlinecite{Glazman90,Matveev91} that the quantization of charge is
not precise: the processes of virtual tunneling between the lead and the
grain result in smearing of charge steps.  The smearing is especially
dramatic in the regime of strong tunneling, when the conductance $G$ of
the tunneling barrier is large, $G\gtrsim e^2/h$. The evolution of the
charge quantization with the increasing $G$ was studied in semiconductor
quantum dots, where the junction conductance can be tuned between $G=0$
and $G\sim e^2/h$.\cite{Westervelt,Molenkamp,Berman} Charge quantization
was also observed\cite{Lafarge} in a metallic ``electron box'' device,
yielding a sharp Coulomb staircase at small $G$. In a metallic device,
one can achieve the tunneling junction conductance $G\gg e^2/h$,
Ref.~\onlinecite{Joyez}. In this case,
theory\cite{Panyukov,Schoen,Grabert} predicts only weak oscillations of
charge, with amplitude $\sim e \exp(-hG/2e^2)$.  However, the layout of a
metallic device does not allow one to vary the junction conductance, and
the crossover between weak and strong tunneling cannot be observed in one
device.

In this paper we study quantum fluctuations of charge of a normal grain
connected by tunnel junction to a superconducting lead. We show that due to
the existence of the gap $\Delta$ in the lead the steps of the Coulomb
staircase remain vertical at any $G$.  The fluctuations result in a finite
slope of the plateaus of the staircase. Unlike the \mbox{N-I-N} case, the
slope may remain small even at $G\gg e^2/h$. One can suppress the gap by
applying a magnetic field,\cite{Lafarge} thus achieving a crossover between
the limits of weak and strong charge fluctuations in one sample.  The main
result of the paper is illustrated by Fig.~\ref{fig:1}, obtained under the
assumption $\Delta\gg E_C$, where $E_C=e^2/2C$ is the charging energy of the
electron box, and $C$ is its total capacitance.

To account for the Coulomb interactions, we include the charging energy term
$(\hat Q-q)^2/2C$ into the Hamiltonian $\hat H$ of the system. The external
charge $q$ here is proportional to the gate voltage. Then  by statistical
averaging of the obvious relation $\partial\hat H/\partial q = (q-\hat Q)/C$,
we find the average charge of the grain
\begin{equation}
  \label{charge}
  Q(q)\equiv\langle Q\rangle = q - C\frac {\partial F(q)}{\partial q}.    
\end{equation}
Here $F(q)=-T\ln Z(q)$ is the free energy.

To find the partition function $Z(q)$, we use the standard effective
action approach by Ambegaokar, Eckern and Sch\"on.\cite{Ambegaokar}  In
this formalism an auxiliary Hubbard-Stratonovich field $\varphi(\tau)$ is
introduced to replace the quadratic in $\hat Q$ interaction term by a
linear one. The electronic degrees of freedom are then traced out,
resulting in the following expression for the partition function:
\begin{equation}
  \label{partition}
  Z(q)=\sum_{m=-\infty}^{+\infty}Z_m e^{i2\pi mq/e},
\end{equation}
with 
\begin{equation}
  \label{action}
  Z_m=\int_{\varphi(0)=0}^{\varphi(\beta)=2\pi m} D\varphi(\tau) 
  \exp\left\{-\int_0^\beta \frac{C\dot\varphi^2}{2e^2} d\tau -
  S_t[\varphi]\right\}. 
\end{equation}
Here the summation over winding numbers $m$ accounts for the discreteness of
charge,\cite{Guinea} and $\beta$ is inverse temperature.\,\,  The tunneling
contribution $S_t[\varphi]$ to the effective 

{\narrowtext
\begin{figure}
\hspace*{1em}
\epsfxsize=.4\textwidth
\epsffile{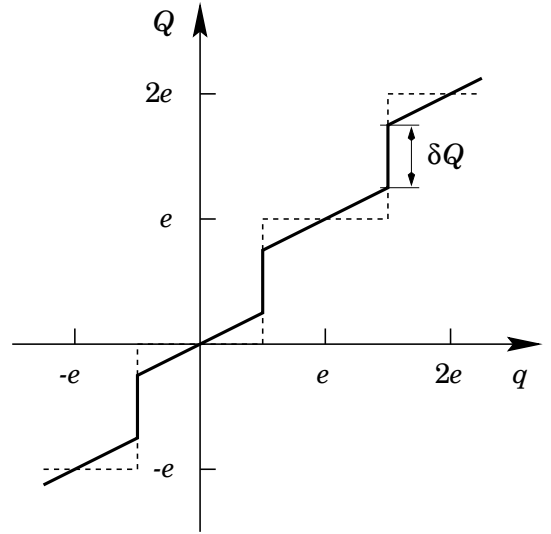}
\vspace{1ex}
\caption{Coulomb staircase for a normal grain connected to a
  superconducting lead. The steps remain vertical at
  any conductance. The plot corresponds to the case $\Delta\gg E_C$, and
  $G=(e^2/\hbar)\Delta/E_C$, see Eq.~(\protect\ref{Qaverage}). Note that
  $G\gg e^2/\hbar$.}
\label{fig:1}
\end{figure}
}
\noindent
action was evaluated for a wide
junction in Ref.~\onlinecite{Ambegaokar}.  It is non-local in time
$\tau$; however, at large $\Delta\gg E_C$ its contribution reduces to a mere
renormalization of capacitance $C$ in the action in Eq.~(\ref{action}).  The
magnitude of the renormalization was found in Ref.~\onlinecite{Ovchinnikov}.
In the case of N-I-S junction the renormalized capacitance is
\begin{equation}
  \label{Ctilde}
  \tilde C = C+\hbar G/2\Delta.
\end{equation}
In this approximation the action in Eq.~(\ref{action}) is quadratic, and
we can easily find the dependence of $Z_m$ on the winding number:
\begin{equation}
  \label{integral}
   Z_m=Z_0\exp\left[-\frac{\tilde C}{2e^2}\frac{(2\pi m)^2}{\beta}\right].
\end{equation}
To study $Z(q)$ at low temperatures $T\ll e^2/\tilde C$, it is convenient
to transform the partition function (\ref{partition}) with the help of the
Poisson summation formula.  Equations (\ref{partition}) and
(\ref{integral}) then yield
\begin{equation}
  \label{Poisson}
  Z(q) = Z_0 \sqrt{\frac{e^2}{2\pi \tilde C T}} 
         \sum_{n=-\infty}^{+\infty} 
             \exp\left[-\frac{(ne-q)^2}{2\tilde C T}\right].
\end{equation}
The largest term in sum (\ref{Poisson}) corresponds to an integer $n$ which is
closest to $q/e$.  At $T\to 0$ this term dominates the sum, and using
Eqs.~(\ref{charge}) and (\ref{Ctilde}), we find
\begin{equation}
  \label{Qaverage}
  Q(q) = e\,{\rm Int }\left(\frac{q}{e}+\frac{1}{2}\right) +
  \frac{q-e\,{\rm Int}\left(\frac{q}{e}+\frac{1}{2}\right)}
{1+\frac{\Delta}{E_C}\frac{e^2}{\hbar G}},
\end{equation}
where ${\rm Int} (x)$ is the integer part of $x$. This is the central result
of the paper; see also Fig.~\ref{fig:1}.   At $G\to0$ the second term in
Eq.~(\ref{Qaverage}) vanishes, leading to a perfect staircase.  At non-zero
$G$ the second term describes the finite slope of the plateaus. According to
Eq.~(\ref{Qaverage}) the steps in $Q(q)$ remain sharp at any conductance, in
contrast to the case of a normal lead.  Even  at $G\sim e^2/\hbar$ the heights
of the steps $e/(1 + E_C\hbar G/\Delta e^2)$ are still close to their nominal
value $e$. 

The fact that $Q(q)$ is discontinuous at half-integer $q/e$ is due to the
density of states in the superconductor vanishing at the Fermi level, and
therefore does not rely on the assumption $\Delta\gg E_C$. Indeed, it is
known\cite{Matveev91} that the problem of charge fluctuations in an electron
box can be mapped onto an effective spin-$\frac12$ Kondo problem.  It follows
from Eq.~(30) of Ref.~\onlinecite{Matveev91} that the height of the charge
step is $\delta Q = 2e\mu$, where $\mu$ is the renormalized value of the spin
of the Kondo impurity.  In the case of a normal metal the spin is completely
screened, $\mu=0$, and the steps are smeared completely, $\delta Q=0$.  If at
least one of the leads is a superconductor, the spin $\mu$ is not comletely
screened, and the step height $\delta Q$ remains finite.

At finite but low temperature $T\ll e^2/\tilde C$, the steps are smeared
slightly. To account for the smearing, one needs to retain two terms, with
$n={\rm Int} (q/e)$ and $n={\rm Int}(q/e)+1$, in the sum (\ref{Poisson}).
In this approximation the charge becomes
\begin{eqnarray}
  \label{finiteT}
  Q(q) &=& e\,{\rm Int}(q/e)+
     \frac{\tilde C - C}{\tilde C}[q-e\,{\rm Int}(q/e)]\
     \nonumber\\
     && + e\,\frac{C}{\tilde C}\left[1+\exp\left(
           \frac{e^2[{\rm Int}(\frac qe)+\frac12-\frac qe]}
                {\tilde C T}\right)\right]^{-1}.
\end{eqnarray}
One can easily check that in the limit of zero temperature this result
reproduces Eq.~(\ref{Qaverage}). 

The effective action technique used here is applicable for tunnel
junctions of wide area, where the conductance is distributed over a large
number $N\gg1$ of transverse modes.  At finite $N$ one may have to take
into account the possibility of two-electron (Andreev) tunneling.
Although the steps in the dependence $Q(q)$ remain vertical, the plateau
conductance due to such processes acquires a correction, which can be
estimated as $\delta Q\sim (\hbar G/e^2)^2 N^{-1}$.  The diffusion of
electrons inside the grain may enhance this correction.  The result can be
expressed in terms of effective number of channels\cite{Hekking} $N_{\rm
  eff}\lesssim N$, which still remains very large due to the smallness of
the Fermi wavelength in metals.

The authors are grateful to I.L. Aleiner for discussions.  The work of
K.A.M. was supported by A.P. Sloan Foundation.  The work of L.I.G. was
supported by NSF Grant DMR-9731756.  The authors also acknowledge the
hospitality of Lorentz Centre at Leiden University where part of the work
was performed.

\end{multicols}


\begin{references}

\bibitem{GrabDev} D.V. Averin and K.K. Likharev, in {\it Mesoscopic
    Phenomena in Solids}, edited by B.L. Altshuler, P.A. Lee, and R.A. Webb
  (Elsevier, Amsterdam, 1991); {\it Single Charge Tunneling,} edited by H.
  Grabert and M.H.  Devoret (Plenum Press, New York, 1992).

\bibitem{Glazman90} L.I. Glazman and K.A. Matveev, Sov. Phys. JETP
  {\bf 71}, 1031 (1990). 

\bibitem{Matveev91} K.A. Matveev, Sov. Phys. JETP {\bf 72}, 892 (1991).

\bibitem{Westervelt}  C. Livermore {\it et al.}, Science, {\bf 274}, 1332
(1996)

\bibitem{Molenkamp}  L.W. Molenkamp, K. Flensberg, and M. Kemerink,
Phys. Rev. Lett. {\bf 75}, 4282 (1995).

\bibitem{Berman} D. Berman {\it et al.}, cond-mat/9803373.

\bibitem{Lafarge} P. Lafarge {\it et al.}, Nature, {\bf 365}, 422, (1993).

\bibitem{Joyez} P. Joyez {\it et al.}, Phys. Rev. Lett. {\bf 79}, 1349
  (1997). 

\bibitem{Panyukov} S.V. Panuykov and A.D. Zaikin, Phys.  Rev. Lett, {\bf
    67}, 3168 (1991).

\bibitem{Schoen} G. Falci, G. Sch\"on, G.T. Zimanyi, Phys.  Rev. Lett,
  {\bf 74}, 3257 (1995).

\bibitem{Grabert}  X. Wang and H. Grabert, Phys. Rev. B, {\bf 53}, 12621
  (1996). 

\bibitem{Ambegaokar} V. Ambegaokar, U. Eckern and G. Sch\"on,
  Phys. Rev. Lett. {\bf 48}, 1745 (1982). 

\bibitem{Guinea} F. Guinea and G. Sch\"on, Europhys. Lett. {\bf 1}, 585
  (1986); G. Sch\"on and A.D. Zaikin, Phys. Rep. {\bf 198}, 237 (1990).

\bibitem{Ovchinnikov} A.I. Larkin and Yu. N. Ovchinnikov, Phys. Rev. B
  {\bf 28}, 6281 (1983). 

\bibitem{Hekking} F.W.J. Hekking and Yu. V. Nazarov, Phys. Rev. B {\bf 49},
6847 (1994).

\end{references}
\end{document}